\documentclass[10pt,letterpaper]{article}
\usepackage{amsmath,amssymb,cite}
\usepackage[colorlinks=false]{hyperref}
\usepackage[pdftex]{color}
\usepackage[pdftex]{graphicx}
\usepackage{changepage}
\usepackage[utf8]{inputenc}
\usepackage{textcomp,marvosym}
\usepackage{fixltx2e}
\usepackage{nameref}
\usepackage{rotating}
\usepackage{microtype}
\usepackage{color}
\usepackage[top=0.75in,left=0.75in,bottom=0.75in,right=0.75in]{geometry}
\usepackage[aboveskip=1pt,labelfont=bf,labelsep=period,justification=raggedright,singlelinecheck=off]{caption}

\makeatletter
\renewcommand{\@biblabel}[1]{\quad#1.}
\makeatother

\date{}

\usepackage{lastpage,fancyhdr,graphicx}
\usepackage{epstopdf}
\newcommand{\pKa}{p$K_a${}}
\newcommand{\isatab}{ISA-Tab}
\newcommand{\isa}{ISA}

\newcommand{\fieldname}[1]{{\texttt{#1}}}

\begin{document}

\begin{flushleft}
{\Large \textbf{An \isatab{} specification for protein titration data exchange}} \\ \mbox{} \\
% Insert Author names, affiliations and corresponding author email.

Chase P.~Dowling$^{1\dagger}$,
Elizabeth Jurrus$^{1\dagger}$,
Sylvia Johnson$^{2}$,
Nathan A.~Baker$^{1,3,\ast}$
\\
\textbf{$^1$} Computational and Statistical Analytics Division, Pacific Northwest National Laboratory, Richland, WA, USA \\
\textbf{$^2$} Advanced Computing, Mathematics, \& Data Division, Pacific Northwest National Laboratory, Richland, WA, USA \\
\textbf{$^3$} Division of Applied Mathematics, Brown University, Providence, RI, USA \\
$\dagger$ These co-authors contributed equally to the manuscript. \\
$\ast$ E-mail: \href{mailto:nathan.baker@pnnl.gov}{nathan.baker@pnnl.gov}
\end{flushleft}

\section*{Abstract}
Data curation presents a challenge to all scientific disciplines to ensure public availability and reproducibility of experimental data.
Standards for data preservation and exchange are central to addressing this challenge: the Investigation-Study-Assay Tabular (\isatab{}) project has developed a widely used template for such standards in biological research.
This paper describes the application of \isatab{} to protein titration data.
Despite the importance of titration experiments for understanding protein structure, stability, and function and for testing computational approaches to protein electrostatics, no such mechanism currently exists for sharing and preserving biomolecular titration data.
We have adapted the \isatab{} template to provide a structured means of supporting experimental structural chemistry data with a particular emphasis on the calculation and measurement of \pKa{} values.
This activity has been performed as part of the broader \textit{\pKa{} Cooperative} effort, leveraging data that has been collected and curated by the Cooperative members.
In this article, we present the details of this specification and its application to a broad range of \pKa{} and electrostatics data obtained for multiple protein systems.
The resulting curated data is publicly available at \url{http://pkacoop.org}.

% \linenumbers

\section*{Introduction}

The preservation and curation of scientific research data has been a topic of discussion since the 1990s \cite{berman00,barrett11}.
In recent years, increased focus has been placed on the standards and storage needs for scientific research.
In most scientific datasets, there are two broad categories of data.
\emph{Ephemeral} data is irreplaceable and cannot be regenerated while \emph{stable} data can be reproduced from source ephemeral data~\cite{gray02}.
The focus of this paper is the metadata associated with the ephemeral data.
Metadata is a common type of data that includes the procedures necessary in order to produce the experimental data, qualitative descriptors, persons and institutions involved, and ontological tags providing semantic information for these data.
Given ephemeral metadata, stable data can be regenerated via computation, simulation, or reproduction of the analysis as prescribed in the metadata.
Ephemeral data, therefore, has intrinsic value for broader meta-analyses of related experimental data sets and reproduction of experimentally observed results~\cite{gray02}.
Therefore, there is a significant need to curate data and metadata.
The careful documentation and storage of ephemeral data (e.g., software used, laboratory parameters, data, etc.) is an important step toward addressing the reproducibility of stable, computational experimental outcomes.

This paper describes the development of a standard to ensure the
preservation and sharing of \pKa{} Cooperative data.  There currently
exist several application-specific data curation efforts that have
resulted in standards that specify how data should be preserved; e.g.,
the Minimum Information About a Microarray Experiment
(MIAME)~\cite{brazma01} and the Systems Biology Markup Language
(SBML)~\cite{thomas13}.  However, given the diversity of experimental and
computational methods used to generate protein \pKa{} data, we wanted to
use a template that was easily extensible, available as open source, and
already used in multiple applications.  The schema we identified as the
most applicable for this study is the Investigation-Study-Assay Tabular
(\isatab{}) format, which is an open-source standard for biological
experimental data~\cite{Sansone2008,ISA-tools}.
\isatab{} was originally  designed for microarray data, but has been
expanded to include a wide range of experimental methods relevant to
protein biophysics \cite{rocca10}, experimental data in biomedicine,
nanomaterials~\cite{thomas13}, and targeted bioinformatics research on
stem cells in the Stem Cell Discovery Engine (SCDE)~\cite{sui12}.  The
SCDE example is relevant to our current work extending \isatab{} to
experimental and computational research.  SCDE is an analytical suite
with a knowledge base that wraps experimental data using the \isatab{}
specification.  This use of the \isatab{} schema demonstrates its
flexibility in configuration, generalized format of data files, as well
as adaptability to biological and chemical data curation uses.

We aim to extend the \isatab{} specification, in a manner similar to the SCDE implementation, for the measurement and calculation of acid dissociation constants (\pKa{s}) in proteins.
The acid dissociation constants of protein residues are relevant to biochemistry and biophysics because they reveal important information about protein energetics, stability, and function.
The importance of accurately capturing these values is discussed at length in the \pKa{} Cooperative article which also describes the origins of our curated data~\cite{nielsen11}.

\section*{Methods} \label{sec:Methods}

\isatab-formatted (meta)data is organized in a directory containing
tab-delimited text files linked by prefixes in the file names
\cite{Sansone2008}.  These text files are divided into three categories:
Investigation, Study, and Assay with the filename prefixes ``i\_'',
``s\_'', and ``a\_'', respectively.  The tab-delimited text files can be
read by a lightweight Java interpreter called the \textit{\isa
creator} or by
a wide range of other parsers.  The \textit{\isa creator} interpreter parses
the \isatab-formatted directory of files based on XML configuration
files.  The configuration file can be modified by direct editing of the
XML or via the \textit{\isa creator {configurator}} software.
The \isatab{} file is static, allowing reuse across multiple datasets.
\textit{{The \isa creator}} and \textit{\isa creator
configurator} tools are open source
tools, easily accessed through their home
page~\cite{ISA-tools}.
We expanded the \isatab{} configuration file by adding additional assay
configurations for assays measuring chemical shift, titration curve
measurements, \pKa{} values, and calculated electrostatic properties such
as free energy.  For the purposes of the \pKa{} Cooperative data
exchange, one journal article is considered to be one investigation and
an individual \isatab{} directory of data files is created for each
article curated.

\begin{figure}[h!]
    \begin{center}
        \includegraphics[width=\linewidth]{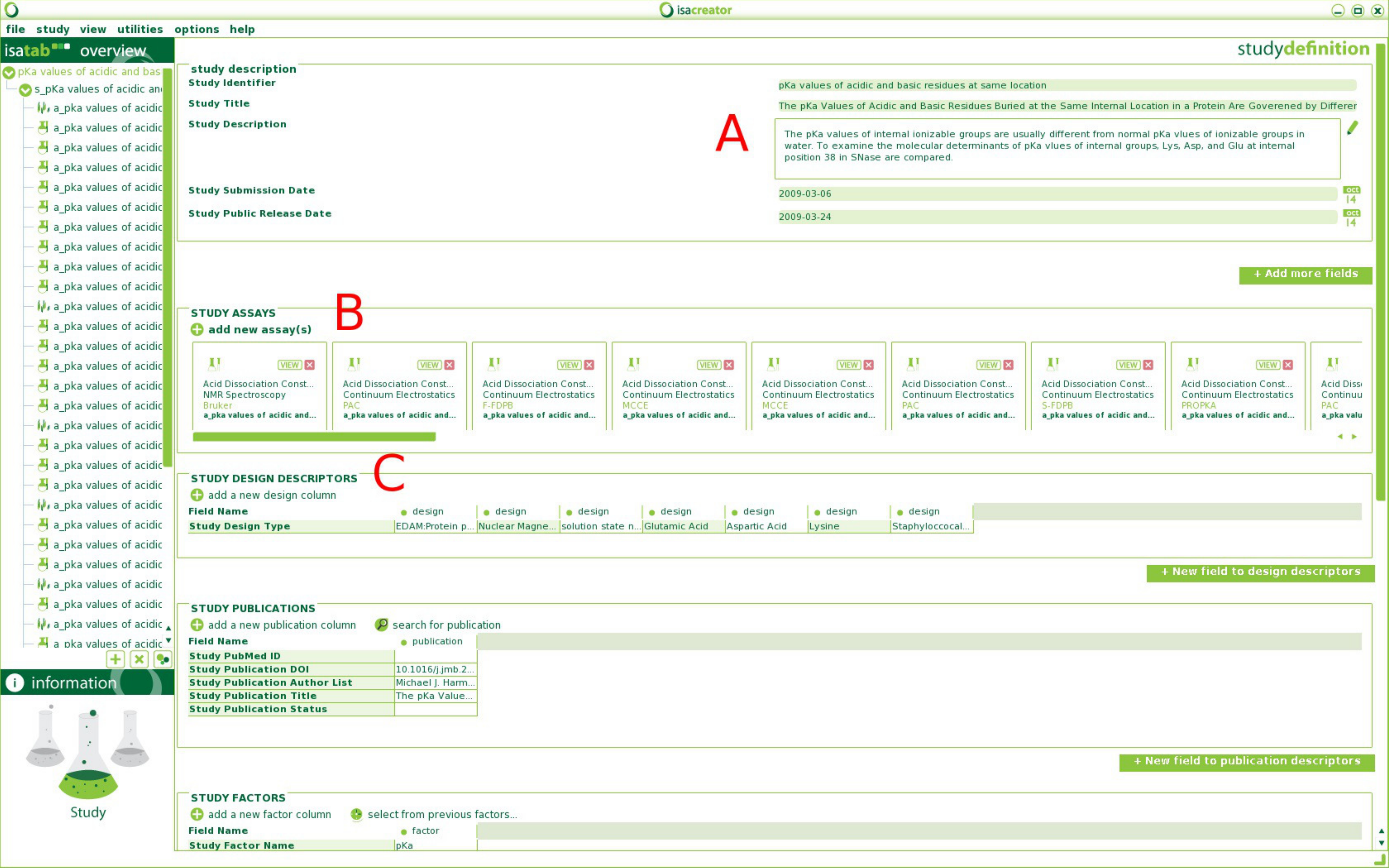}
    \end{center}
    \caption{{\bf \textit{\isa creator} view of investigation file}.
    (A) titles of the studies, researcher contact information, a summary of work, and date of publication; (B) pointers to the various study assay files in the \fieldname{Study Assay} section; and (C) study design descriptors from several ontologies.}
    \label{fig.investigation}
\end{figure}

Our customized \isatab{} configuration was tested against a published journal article~\cite{harms09} from the Garc\'ia-Moreno lab that uses both NMR spectroscopy and continuum electrostatics methods to measure/calculate \pKa{} values for internal residues of a hyperstable variant of staphylococcal nuclease.
An \textit{\isa creator} view of the investigation file is shown in Figure~\ref{fig.investigation} displaying information extracted from the Gar\'ia-Moreno journal article.
The investigation file contains several important elements:  (A) titles of the studies, researcher contact information, a summary of work, and date of publication; (B) pointers to the various study assay files in the \fieldname{Study Assay} section; and (C) study design descriptors from several ontologies, including the Ontology for Biomedical Investigations (OBI) \cite{Brinkman2010}, the National Center for Biotechnology Information (NCBI) Taxonomy \cite{Benson2009}, and the National Cancer Institute (NCI) Thesaurus \cite{DeCoronado2009}.
The left hand column in Figure~\ref{fig.investigation} displays a list of all study and assay files associated with the investigation.

\begin{figure}[h!]
    \begin{center}
        \includegraphics[width=\linewidth]{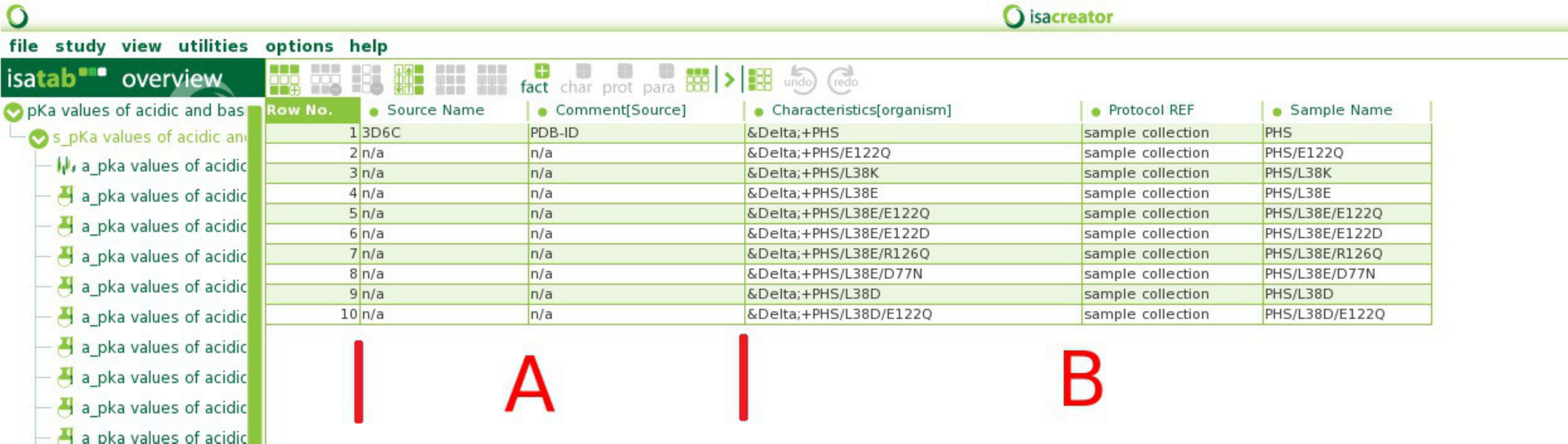}
    \end{center}
    \caption{{\bf ISA-Tab view of study file}.  (A) a reference molecule ID (if available) for the protein under study and (B) a list of a list of all proteins and variants examined in the study.}
    \label{fig.study}
\end{figure}

An investigation generally includes several studies; individual study
files can be viewed in \textit{\isa creator} by selecting items from the lefthand side of the screen shown in Figure~\ref{fig.investigation}.
An example \textit{\isa creator} study file view is shown in Figure~\ref{fig.study} and includes (A) a reference molecule ID (if available) for the protein under study and (B) a list of a list of all proteins and variants examined in the study.
In the current example, the PDB ID is used as the molecule reference in the \fieldname{Source Name} field; however, other IDs (GenBank, etc.) could be used as well with the appropriate references added to the \fieldname{Comment[Source]} field.
Mutants are described by the \fieldname{Characteristics[organism]} field and given unique identifiers in the \fieldname{Sample Name} field.
The left hand column in Figure~\ref{fig.study} displays a list of assay files associated with the study.

\begin{figure}[h!]
    \begin{center}
        \includegraphics[width=\linewidth]{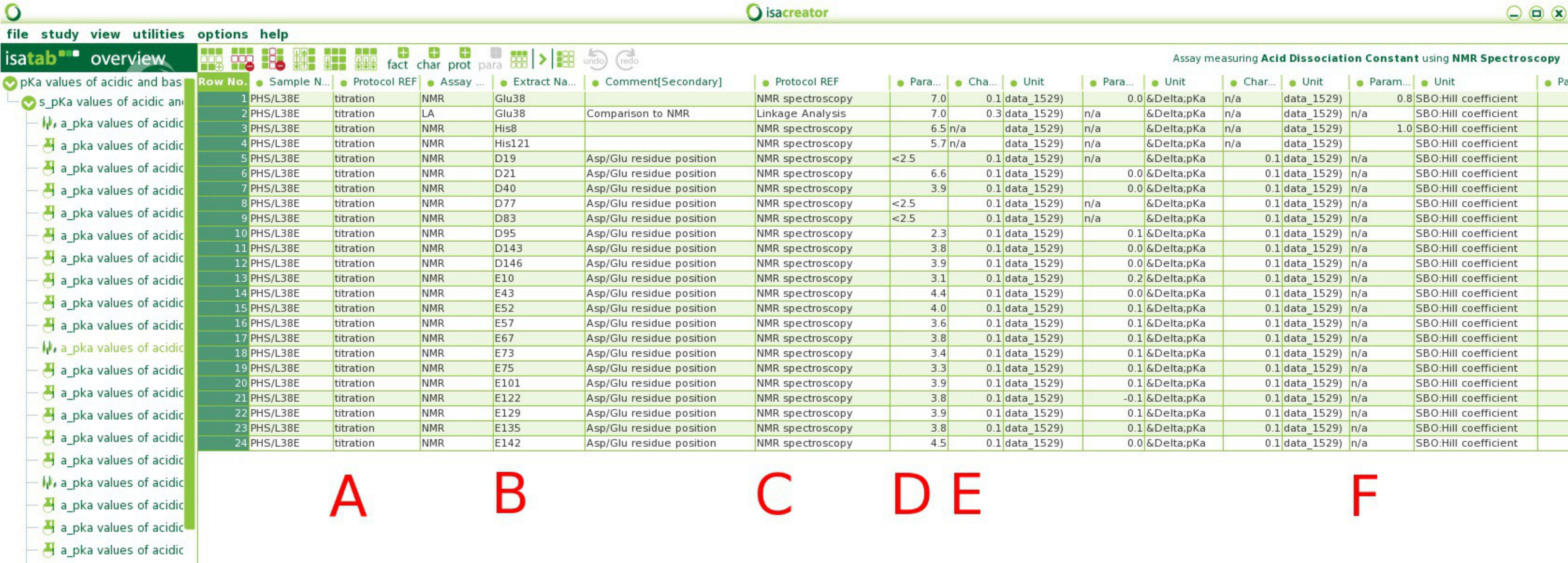}
    \end{center}
    \caption{{\bf ISA-Tab view of assay file}. (A) A single measurement technology or software simulation outcome for (B) a single protein variant identified in the parent study file using (C) a specific measurement protocol to obtain (D) a \pKa{} value and (E) standard deviation as well as (F) other derived values.}
    \label{fig.assay}
\end{figure}

A study generally includes several assays; individual assay files can be
view in \textit{\isa creator} by selecting items from the lefthand side of the screen show in Figure~\ref{fig.study}.
An example \textit{\isa creator} assay file view is shown in Figure~\ref{fig.assay} and includes (A) a single measurement technology or software simulation outcome for (B) a single protein variant identified in the parent study file using (C) a specific measurement protocol to obtain (D) a \pKa{} value and (E) standard deviation as well as (F) other derived values.
A number of ephemeral data attributes are exposed for this assay, including the protein residue (available in the \fieldname{Extract Name} and the \fieldname{Protocol REF} fields.
The \fieldname{Protocol REF} field includes an ontology concept to reference the experimental (or computational) approach used for each measurement as well as methods such as linkage analysis used to derive \pKa{} values from the data.
Several protocols have been implemented into our configuration file, including both computational and experimental methods (described in the Results section below).
In addition to capturing as much ephemeral non-reproducible data as possible; e.g., additional fitting parameters used in data analysis in \fieldname{Parameter Value[Derived Experimental Data]} and associated fields such as linkage (or Hill) coefficients.

\section*{Results and discussion}

We tested our extended \isatab{} data-sharing format by curating information from 20 published manuscripts with protein titration data.
These manuscripts included NMR spectroscopy \cite{Cocco1992, Bartik1994, Assadi-Porter1995, Kesvatera1996, Tan1995, Chiang1996, Gooley1998, Koide2001, Lund-Katz2001, Garcia-Mayoral2003, Sakurai2007, Harms2008, Castaneda2009, harms09, Webb2011, Fitch2002}, UV circular dichroism \cite{Isom2010a, Isom2011}, and continuum electrostatics calculations \cite{Kesvatera1996, Gooley1998, Garcia-Mayoral2003, Harms2008, Castaneda2009, harms09, Zheng2009, Webb2011, Isom2011, Meyer2011, Fitch2002}.
To increase the diversity of the test dataset and test generalizability beyond \pKa{} values, we also included electrochemistry data of heme proteins \cite{Zheng2009}.
The curation of both experimental and computational data is an important aspect of this work.
The \pKa{} Cooperative recently led a blind prediction challenges for \pKa{} data given small amounts of seed experimental information~\cite{olsson11, alexov11}.
Meaningful analysis and comparison \cite{nielsen11, gosink13} between computational predictions requires a standardized data exchange format that captures relevant metadata.
The curated \isatab{}-formatted versions of these articles as well as the \isatab{} configuration files can be downloaded from (\url{https://pkacoop.org}).
{The Supporting Information for this manuscript contains a ZIP file of the \isatab{} materials as well as step-by-step instructions for using the files with the \textit{\isa creator} software.}
This curated data is managed via GitHub due to its open availability and participant-sourced revision process.

In future work, we hope to expand the set of curated data and supported assays as well as extend curation to include \pKa{}s measured in small molecule compounds such as pharmaceuticals~\cite{babic07}.
Future work will also include automatic formatting of CSV spreadsheets from existing data tables into an \isatab{} compliant format to decrease curation time
requirements for current and future research.
As minimal requirements for \pKa{} data sharing, we recommend that the \pKa{} parameter values, statistical characteristics of measurements and fits (e.g., standard errors), fitting models, assay types, residues measured or continuum electrostatics model used be included in reported data.
We hope that the \isatab{} \pKa{} data sharing format will enjoy adoption within and beyond the \pKa{} Cooperative and serve as a starting point for broader adoption of informatics standards by the molecular biophysics community.

\clearpage

% \nolinenumbers

\section*{Acknowledgments}
The authors gratefully acknowledge support from OpenEye Scientific Software and NIH grant R01 GM069702.
The authors thank Dr.~Susanna-Assunta Sansone, University of Oxford e-Research Centre, for helpful discussions and support of the \isatab{} standard.
We would also like to thank the \pKa{} Cooperative for their contributions to this effort.

% \nolinenumbers

\bibliography{isatab-pka-specification}

\section*{Supporting information}

{The \isatab-formatted data, described above, are available for download from the \pKa{} Co-operative GitHub web page: \url{http://pkacoop.org}.
A ZIP format is also provided as Supporting Information to this manuscript.
The ZIP archive includes the \isatab{} configuration file described in this paper as well as the individual datasets.
The \textit{\isa creator} application full version is required to view the \isatab{} files through the customized user interface rather than as spreadsheets; this software is freely available from \url{http://www.isa-tools.org/software-suite/}.
After creating an account in \textit{\isa creator}, users can add the \texttt{isaconfigPChem} directory to the list of available \textit{\isa creator} configurations.
Alternatively, users can copy this directory to their \textit{\isa creator} installation directory (under Configurations) so that it will be available by default.
The provided \pKa{} Co-operative \isatab{} (found in the \texttt{isa-tab-data directory} of Supporting Information) can be loaded through the main menu of the \textit{\isa creator} software.}

\end{document}